\documentclass{aa}  

\usepackage{graphicx}
\usepackage{bm}
\usepackage{natbib}
\usepackage{txfonts}
\begin{document} 

\title{The periodic origin of fast radio bursts}


\author{Yu-Jia Wei\inst{1}
        \and Zhen-Yin Zhao\inst{1}
        \and Fa-Yin Wang\inst{1,2}
}

\institute{School of Astronomy \& Space Science, Nanjing University, Nanjing 210023, China \\
        \email{fayinwang@nju.edu.cn}
        \and
        Key Laboratory of Modern Astronomy and Astrophysics (Nanjing University), Ministry of Education, Nanjing 210093, China\\
}


\abstract
{Fast radio bursts (FRBs) are pulsed radio signals with a duration of milliseconds and a large dispersion measure (DM). Recent observations indicate that FRB 180916 and FRB 121102 show periodic activities. Some theoretical models have been proposed to explain periodic FRBs, and here we test these using corresponding X-ray and $\gamma$-ray observations. We find that the orbital periodic model, the free precession model, the radiation-driven precession model, the fall-back disk precession model where eccentricity is due to the internal magnetic field,  and the rotation periodic model are not consistent with observations. The geodetic precession model is the most likely periodic model for FRB 180916. We also propose methods to test the periodic models with yet-to-be-obtained observational data in the future.}

\keywords{Stars: magnetars -- Stars: neutron -- fast radio bursts}

\maketitle

%
\section{Introduction}           
\label{sect:intro}

Fast radio bursts (FRBs) are transient radio signals with a duration of milliseconds (ms), flux of $\sim 0.1-1 \ \rm Jy,$ and a large dispersion measure (DM; \citealt{Cordes_2019,Petroff_2019,ZhangBing_2020,XiaoDi_2021}). The first recorded FRB, named the "Lorimer burst" (also known as FRB 010724), was discovered in 2007 (\citealt{Lorimer_2007}) after this signal was first detected by the Parkes 64-meter telescope in Australia on July 24, 2001. Recently, there have been two intriguing discoveries. One is that a period of $16.35 \ {\rm d}$ (\citealt{2020Natur.582..351C}) and a period of $157 \ {\rm d}$ (\citealt{Rajwade_2020,Cruces_2021}) were found in FRB 180916 and FRB 121102, respectively. The other is that FRB 200428 was detected (\citealt{Bochenek_2020, 2020Nature..587...54}) and was related to a hard X-ray burst originating from SGR 1935+2154 in the Milky Way (\citealt{Li_2021, Mereghetti_2020, Ridnaia_2021, Tavani_2020_xray}). This suggests that at least a portion of FRBs are produced by magnetars.
Some theoretical models based on magnetars have been proposed (\citealt{Kulkarni_2014,Lyubarsky_2014,Katz_2016,Murase_2016,Beloborodov_2020,Metzger_2019,Wang_2020}). Interestingly, the statistical properties of the repeating
bursts are consistent with Galactic magnetar bursts (\citealt{Wang_2017,Wadiasingh_2019,Cheng_2020,ZhangG2021,Wei2021,Sang2021}).

The first repeating FRB, FRB 121102 \citep{Spitler2016}, was found to be in an extreme magnetized environment \citep{2018NatureMichilliAnextreme}, which was revealed by the very high rotation measure (RM$\sim 10^5$ rad m$^{-2}$). The RM decreases by $\sim 34 \%$ in 2.6 yr \citep{Hilmarsson2021}. The high and rapidly variable RM can be well understood in the magnetar nebula model (\citealt{Margalit2018,Katz2021MNRAS.501L..76K}). An increase in the DM of FRB 121102 was found from long-term observations \citep{Hessels2019,Josephy2019,Oostrum2020,Li2021}. The DM and RM evolution of FRB 121102 can be associated with the expanding magnetar wind nebula and shocked shell in a compact binary merger scenario \citep{zhao2021}. However, such abnormal variations of DM and RM seem absent for other FRBs. The DM of FRB 180916 is almost constant ($\Delta \mathrm{DM} < 0.1$ pc cm$^{-3}$, \citealt{2020Natur.582..351C,Nimmo2021,Pastor-Marazuela2021}).
\cite{Pleunis_2021} found that the RM of FRB 180916 is very small ($\rm RM \approx 115 \ rad \ m^{-2}$) and shows very small variations ($\Delta \mathrm{RM} \sim 2-3$ rad m$^{-2}$). The small and stable RM implies that the environmental magnetic field strength is much lower than that of FRB 121102. From the upper limits of X-ray and $\gamma$-ray emission, \cite{Tavani_2020} gave constraints on the dissipation of magnetic energy of FRB 180916, which is inconsistent with the model involving strong magnetic field. 

Using the simultaneous observations of FRB 180916 from the Apertif receiver (1220 MHz and 1520 MHz) on the Westerbork Synthesis Radio Telescope and the Low Frequency Array (LOFAR, 110 MHz and 190 MHz), \cite{Pastor-Marazuela2021} found that the active window of FRB 180916 is narrower and earlier at higher frequencies. The activity window at 150 MHz is narrower than those at 600 MHz and 1.4 GHz, and its peak activity is about 0.7 days earlier than that observed by CHIME/FRB in 600 MHz. The full width at half height of the activity of FRB 180916 observed by Apertif is 1.1 days, while that observed by CHIME/FRB is 2.7 days. Furthermore, the peak of the active period observed by LOFAR seems to arrive about 2 days later than that observed by CHIME/FRB, but their low number of detection prevents them from obtaining a better estimate of the active window. 

The source of FRB 121102 is assumed to be a young magnetar with an age of $\sim 10-40$ yr \citep{Margalit2018}. \cite{Marcote_2020} speculated that the age of the progenitor of FRB 180916  is likely to be about 300 years based on the same RM decay model of FRB 121102. From the distance to the nearest young stellar clump, \cite{Tendulkar2021} inferred an age for the progenitor for FRB 180916 of between 800 kyr and 7 Myr, which seems to be inconsistent with the young active magnetars. But if the magnetar was born in the compact binary mergers, the observed spatial offset is also possible. In this work, we follow the young magnetar models and the age speculation from \cite{Marcote_2020}.

To explain the periodic activity, several types of models were proposed, including the collisions of pulsars and extra-galactic asteroid belts (\citealt{Dai_2020_ApJ...895L...1D}), the orbital models (\citealt{Ioka_2020, Lyutikov_2020, Wada_2021_arXiv210514480W,LiQC_2021}), the precession models (\citealt{Levin_2020, Sob_yanin_2020, Tong_2020, Yang_Zou_2020, Zanazzi_2020,Chen2021}) and the rotation period model (\citealt{Beniamini_2020,Xu_2021}). Recently, \cite{Katz_2021} tested some types of period models based on observations by \cite{Pastor-Marazuela2021} and \cite{Pleunis_2021}. Little work has been carried out to evaluate these period models, especially those based on the latest observations, which are discussed below.

In this work, we provide constraints on these models for FRB 180916  using recent observations (\citealt{Marcote_2020, Pastor-Marazuela2021, Pleunis_2021, Tavani_2020}). In addition, to explain the 16-day period, the predictions of these models should also be consistent with the stable DM and RM, the frequency-dependent active window, the weak internal magnetic field, and the age of about 300 years. Because of the limitations of the observations, we only tested the orbital models, precession models, and rotational models. 
We find that the geodetic precession model is the most likely model for FRB 180916. 

The rest of this paper is organized as follows. Section~\ref{sect:obser} shows the observations used in this paper. In Sect.~\ref{sect:constr}, we constrain several types of period models using these recent observations. Finally, we provide  a discussion of our results and a summary in Sect.~\ref{sect:sum}. 

\section{Recent observational results}
\label{sect:obser}

Four observations are used in this paper to constrain periodic models.
The first observation is that \cite{Pastor-Marazuela2021}, using simultaneous observations of FRB 180916 from the Apertif and LOFAR, found that the active window of FRB 180916 is narrower and earlier at a higher frequency. In addition, the dispersion measure (DM) is constant in the observations at these frequencies, and the maximum signal-to-noise DM is $349.00\ \rm pc \ cm^{-3}$ in the LOFAR observations. Furthermore, \cite{Pleunis_2021} found that the value of RM of FRB 180916 is very small, $\rm RM \approx 115 \ rad \ m^{-2}$.

The second observation is based on the observation of FRB 180916 (\citealt{Pastor-Marazuela2021, Pleunis_2021}); \cite{Katz_2021} proposed a limitation on the variation of angular velocity $\dot{\Omega}$:
\begin{equation}\label{Eq:constrain2}
        |\dot{\Omega}| = \frac{8 |\Delta \phi|}{T^2} \lesssim 5 \times 10^{-16} \ \rm s^{-2},
\end{equation}
where $\Delta \phi$ is the phase deviation from the exact periodic endpoint in a period fitted by the midpoint of the data, $T = 49 P \approx 7 \times 10^7 \ \rm s$ is the observation duration, and $P$ is the observation period. It is assumed that the angular velocity $\Omega$ is fitted to the data from the whole data interval $T$ and \cite{Katz_2021} proposed $|\Delta \phi| \lesssim 0.05 {\ \rm cycle} \approx 0.3 \ \rm radian$, which is estimated from the observation of \citet{Pastor-Marazuela2021} and \citet{Pleunis_2021}. Also, \cite{Katz_2021} mentions that Eq.~(\ref{Eq:constrain2}) will rapidly become more stringent as $T$ increases, or that a significant non-zero rate of angular velocity change $\dot{\Omega}$ will be observed.

\begin{figure}
	\resizebox{\hsize}{!}{\includegraphics{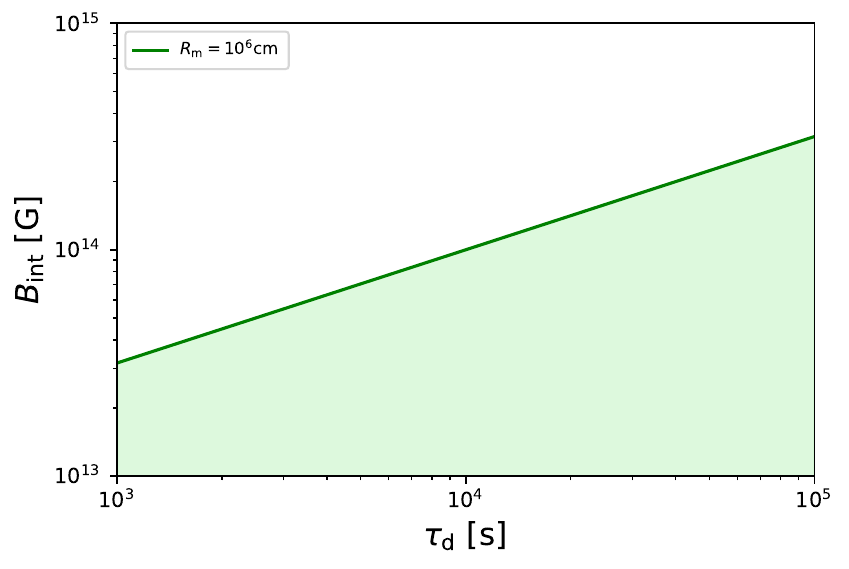}}
    \caption{Constraint on the internal magnetic field $B_{\rm int}$ according to Eq.~(\ref{Eq:constrain2}). The green line shows the condition of $R_{\rm m} = 10^6 \ \rm cm$. The colored area represents the allowed range of the internal magnetic field when the value of $\tau_{\rm d}$ is from $10^3 \ \rm s$ to $10^5 \ \rm s$.} 
    \label{Fig:constrain3}
\end{figure}

The third observation is that based on the X-ray and $\gamma$-ray observation results of FRB 180916 obtained by Astro-Rivelatore Gamma a Immagini Leggero (AGILE) and Swift: \cite{Tavani_2020} derived the most stringent constraints so far which are applicable to magnetar-type neutron stars:
\begin{equation}\label{Eq:constrain3}
        R_{\rm m, 6}^3 B_{\rm int, 16}^2 \tau_{\rm d, 8}^{-1} \lesssim 1,
\end{equation}
where $R_{\rm m}$ is the magnetospheric radius of the magnetar, $B_{\rm int}$ is the internal magnetic field intensity, and $\tau_{\rm d}$ is the ambipolar diffusion timescale of the internal magnetic field. In addition, limited by the observation time, the magnetic dissipation timescale $\tau_{\rm d}$ is $\sim 10^3 - 10^5 \ \rm s$. Therefore, if we set $R_{\rm m}$ to be $10^6 \ \rm cm$, the value of $B_{\rm int}$ should be $\sim 10^{13.5} - 10^{14.5} \ \rm G$, as shown in Fig.~\ref{Fig:constrain3}.

The fourth observation is that \cite{Marcote_2020} estimated the age of the progenitor of FRB 180916  to be about 300 years. Therefore, assuming that the
progenitor of FRB 180916 is a magnetar and combining the relation between the age $t$ and the rotation period $P_{\rm rot}$ of magnetar (\citealt{Levin_2020}):
\begin{equation}
        t \approx 7.5 \ {\rm yr} \ P_{\rm rot, 0}^2 B_{\rm dip, 15}^{-2},
\end{equation}
we can get the rotation period $P_{\rm rot} = 2 \ \rm s$ for the progenitor.

\section{Constraints on periodic models}
\label{sect:constr}

In this section, we use the observations described in Sect.~\ref{sect:obser} to constrain the periodic models for FRB 180916. We mainly limit ourselves to three types of periodic model:
\begin{enumerate}
                \item Orbital model: In a binary system, FRBs are generated by a neutron star. This radio wave might therefore be obscured by or interact with the wind of the companion star, which could cause the periodicity (\citealt{Ioka_2020, Lyutikov_2020}).
                \item Precession models: The precession might be caused by the free precession of a magnetically twisted nonspherical neutron star (\citealt{Levin_2020, Zanazzi_2020}) or the forced precession which might be caused by the anomalous electromagnetic moment (\citealt{Sob_yanin_2020}) or the fall-back disk (\citealt{Tong_2020}), or the orbital induced spin precession of the neutron star in a binary system (\citealt{Yang_Zou_2020}). For convenience, we refer to these as the free precession model, the radiation-driven precession model, the fall-back disk precession model, and  the geodetic precession model, respectively.
                \item Rotation model: the observational period is regarded as the rotation period of a neutron star (\citealt{Beniamini_2020,Xu_2021}).
\end{enumerate}

Additionally, in the following sections, we use the sign $A_x = A/10^x$ in the cgs unit. Unless specifying otherwise, in the following calculations we regard the mass of the neutron star to be $M_{\rm ns} = 1.4 \ M_{\odot}$ and the radius of the neutron star to be $R_{\rm ns} = 10^6 \ \rm cm$.

\subsection{Constraints on the orbital period model}

According to the first observation introduced in Sect.~\ref{sect:obser}, the DM should be a constant, which is consistent with the orbital period model. In addition, according to this observation, the RM is very small ($\sim 100 \ \rm rad \ m^{-2}$) and the active window should be narrower at higher frequencies. However, for the orbital period model, the predicted RM is very large ($\sim 10^3 \ \rm rad \ m^{-2}$) and the active window will be wider on a higher frequency. Therefore, this model cannot explain the periodicity of FRB 180916. \cite{Wada_2021_arXiv210514480W} proposed two possible scenarios to explain this phenomenon, but these two scenarios need extremely strict conditions.

\subsection{Constraints on the precession model}\label{sect:constr_2}

In this section, we mainly use the second and third observations to constrain these period models, except the orbital period model.

\subsubsection{Constraints on the free precession model}

For the free precession model, the derivation of the period of the precession $\dot{P}_{\rm pre}$ is
\begin{equation}\label{Eq:FP_dotPpre}
                \dot{P}_{\rm pre} = \frac{P_{\rm pre}}{2t},
\end{equation}
where $P_{\rm pre}$ is the precession period and $t$ is the age of the neutron star. It should be noted that in \cite{Levin_2020}, the actual age $t$ of the neutron star is used to calculate the precession period $P_{\rm pre}$, but \cite{Katz_2021} proposed that the spin-down timescale $t_{\rm sd}$ may be more appropriate because the neutron star may be born when its spin period and precession period are close to the current value. The $t_{\rm sd}$ is derived as
\begin{equation}\label{Eq:FP_tsd}
                t_{\rm sd} = \frac{12 c^3 M_{\rm ns}}{5 B_{\rm dip}^2 R_{\rm ns}^4 \Omega_{\rm rot}^2} \approx 144.6 \ {\rm yr} \ B_{\rm dip,15}^{-2} R_{\rm ns,6}^{-4} {(\frac{M_{\rm ns}}{1.4 \ M_{\odot}})} P_{\rm rot,0}^2,
\end{equation}
where $B_{\rm dip}$ is the dipole field and $\Omega_{\rm rot} = 2 \pi / P_{\rm rot}$ is the spin angular frequency.
        
Combining this latter equation with Eq.~(\ref{Eq:FP_dotPpre}), we can obtain the change rate of precession angular velocity:
\begin{equation}
        \dot{\Omega}_{\rm pre} = - \frac{\Omega_{\rm pre}}{2 t}.
\end{equation}
Combining Eq.~(\ref{Eq:constrain2}) with $|\Delta \phi| \lesssim 0.3 \ \rm radian$ and $T \approx 7 \times 10^7 \ \rm s$, we can rewrite the above equation as
\begin{equation}\label{Eq:constr2_FP_t}
        t = \frac{\pi T^2}{8 P_{\rm pre} |\Delta \phi|} \gtrsim 150 \ \rm yr.
\end{equation}
In this way, we can obtain a lower limit on the age of the neutron star which is consistent with the 300-year age of the magnetar of FRB 180916 speculated by \cite{Marcote_2020}. This lower limit might increase rapidly as the observation time increases because it is highly dependent on $T$. In this way, this model can be thoroughly tested in future. 

Also, \cite{Zanazzi_2020} gave the relationship between rotation period $P_{\rm rot}$ and precession period $P_{\rm pre}$ of the magnetar (also see in \citealt{Levin_2020}):
\begin{equation}\label{Eq:FP_1}
        P_{\rm pre} = \frac{P_{\rm rot}}{\epsilon {\rm cos} \theta},
\end{equation}
where $\epsilon$ is the ellipticity and $\theta$ is the angle between the rotation angular velocity vector $\bm{\Omega}$ and the principle axis with the greatest moment of inertia, which we call the $z$-axis in the following. \cite{Levin_2020} set the value of $\theta$ to $0$. \cite{Zanazzi_2020} considered five factors that may contribute to the ellipticity $\epsilon$ but finally decided that just four of them are possible.

If we use the spin-down age $t_{\rm sd}$ of the neutron star to replace $t$, we can get the dipole field $B_{\rm dip}$ by combining Eq.~(\ref{Eq:constr2_FP_t}) with Eq.~(\ref{Eq:FP_1}) and Eq.~(\ref{Eq:FP_tsd}):
\begin{equation}\label{Eq:constr_FP_Bdip}
        B_{\rm dip} = \epsilon {\rm cos} \theta \sqrt{\frac{3 c^3 M_{\rm ns} P_{\rm pre}^3 |\dot{\Omega}_{\rm pre}|}{5 \pi^3 R_{\rm ns}^4}} = \epsilon {\rm cos} \theta \sqrt{\frac{24 c^3 M_{\rm ns} P_{\rm pre}^3 |\Delta \phi|}{5 \pi^3 R_{\rm ns}^4 T^2}}.
\end{equation}

\begin{figure}
	\resizebox{\hsize}{!}{\includegraphics{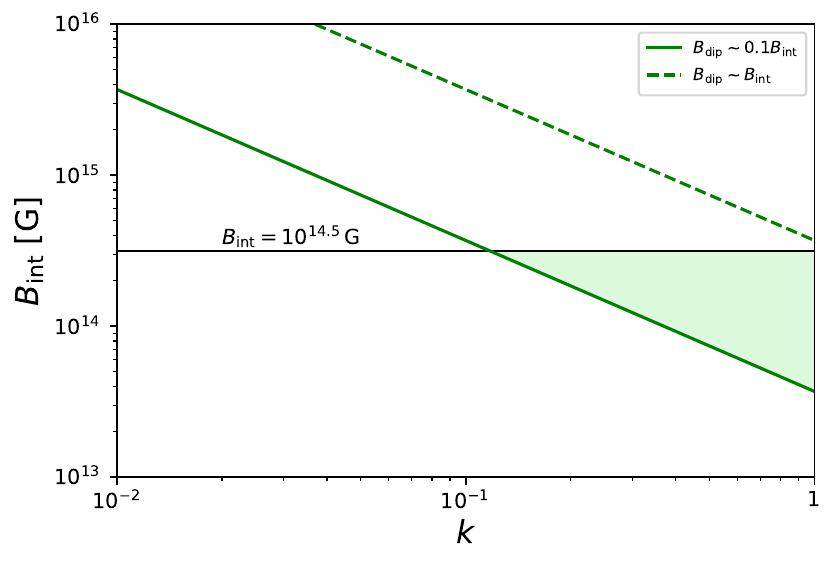}}
    \caption{Relation between the internal magnetic field $B_{\rm int}$ and the numerical coefficient $k$ is drawn according to Eq.~(\ref{Eq:constr_FP_Bdip}). The black horizontal line represents the maximum value of the internal magnetic field limited by Eq.~(\ref{Eq:constrain3}) when the magnetosphere radius $R_{\rm m} = 10^6\ \rm cm$. In addition, the green lines show the relationship between $B_{\rm int}$ and $k$ drawn from Eq.~(\ref{Eq:constr_FP_Bdip}). We show two different conditions, $B_{\rm dip} \sim 0.1 B_{\rm int}$ and $B_{\rm dip} \sim B_{\rm int}$. The colored area shows the allowable region. } 
    \label{Fig:constr_FP_Bint-k}
\end{figure}

Firstly, for the ellipticity caused by the internal magnetic field, $\epsilon = \epsilon_{\rm mag}$, the internal magnetic field $B_{\rm int}$ gives
\begin{equation}\label{Eq:FP_emag}
        \epsilon_{\rm mag} = \beta \frac{R_{\rm ns}^4 B_{\rm int}^2}{G_0 M_{\rm ns}^2} \approx k \times 10^{-4} B_{\rm int,16}^2 R_{\rm ns, 6}^4 {(\frac{M_{\rm ns}}{1.4 \ M_{\odot}})}^{-2},
\end{equation}
where $G_0$ is the gravitational constant, and $R_{\rm ns}$ and $M_{\rm ns}$ are the radius and the mass of the magnetar, which we set to be the typical values of the neutron star. $\beta$ and $k$ are the numerical coefficients which should satisfy $|\beta| \leq 1$ and $k \leq 1$, and \cite{Levin_2020} thought that $k \ll 1$ in this condition. Using the magnetic field value of $B_{\rm dip} \sim 0.1 B_{\rm int}$ in the shock maser model of FRB, we can derive the limitation of $B_{\rm int}$ with Eq.~(\ref{Eq:constr_FP_Bdip}):
\begin{equation}
        \begin{aligned}
                & B_{\rm int} = \frac{G_0 T}{10 k R_{\rm ns}^2 {\rm cos}\theta} \sqrt{\frac{5 \pi^3 M_{\rm ns}^3}{24 c^3 P_{\rm pre}^3 |\Delta \phi|}} \\
                & B_{\rm int} \gtrsim 3.7 \times 10^{13} \ {\rm G} \  k^{-1} ({\rm cos} \theta)^{-1},
        \end{aligned}
\end{equation}
where $k \ll 1$. This equation gives a lower limit on the internal field. In this equation, if $k = 0.12$ is assumed and $\rm cos \theta$ takes the maximum value which is $1$, we can obtain the limit of the internal magnetic field, that is $\gtrsim 3 \times 10^{14} \ \rm G,$ which is consistent with the third observation in Sect.~\ref{sect:obser} (i.e., that the magnitude of $B_{\rm int}$ should be $\sim 10^{13.5} - 10^{14.5} \ \rm G$). In addition to adjusting the value of $k$, $B_{\rm dip} \sim 0.1 B_{\rm int}$ is also uncertain, and the predicted values of this model can be adjusted to conform to the observations, as shown in Fig.~\ref{Fig:constr_FP_Bint-k}.

It is important to note that equation 7 in \cite{Katz_2021} contains a typo, where the exponent of $k$ should be $1$ instead of $2$. The correct equation should be
\begin{equation}
        B \sim \frac{10^{36} \ {\rm G^2}}{k} \sqrt{\frac{ {\rm sin}^2 \theta \ R^6 \pi^3 T^2}{6I c^3 P_{\rm pre}^3 \Delta\phi}} = 5 \times 10^{14}\ {\rm G} \ \frac{{\rm sin} \theta}{k} \sqrt{\frac{0.3}{\Delta\phi}} \frac{T}{1 \ {\rm yr}}.
\end{equation}
Also, \cite{Katz_2021} used the approximation of $B_{\rm int} \sim B_{\rm dip}$, and in this paper we use $B_{\rm dip} \sim 0.1 B_{\rm int}$. Furthermore, we also use other different relationships to get the limit of the magnetic field. In this way, the condition that the limitation in this paper is different from that of \cite{Katz_2021} can be understood. And these two limitations can be confirmed by each other.

Secondly, for the ellipticity caused by the near-region dipole field associated with neutron stars, its ellipticity is
\begin{equation}\label{Eq:FP_edip}
        \epsilon_{\rm dip} = \frac{3B_{\rm dip}^2 R_{\rm ns}^5}{20 I c^2} \approx 1.5 \times 10^{-7} B_{\rm dip,15}^2 R_{\rm ns,6}^3 {(\frac{M_{\rm ns}}{1.4 \ M_{\odot}})}^{-1},
\end{equation}
where $I = \frac{2}{5} M_{\rm ns} R_{\rm ns}^2$ is the rotational inertia of the magnetar. We can obtain the limitation of this dipole field with Eq.~(\ref{Eq:FP_edip}) and Eq.~(\ref{Eq:constr_FP_Bdip}):
\begin{equation}
        B_{\rm dip} = \frac{4 T}{3 R_{\rm ns} {\rm cos} \theta} \sqrt{\frac{5 \pi^3 c M_{\rm ns}}{6 P_{\rm pre}^3 |\Delta \phi|}} \gtrsim 4.7 \times 10^{15}  \ {\rm G} \ {({\rm cos} \theta)}^{-1}.
\end{equation}
This equation gives a lower limit of the dipole magnetic field, and even if the ${({\rm cos} \theta)}^{-1}$ takes the minimum value of $1$, it will exceed the magnetic field range of the observed limit.

Thirdly, for the ellipticity caused by the two components $B_{\parallel}$ and $B_{\delta}$ of the quadrupole magnetic field associated with the neutron star,  $\epsilon \sim \epsilon_{\parallel}$ and $\epsilon \sim \epsilon_{\delta}$. According to \cite{Zanazzi_2020}, we have
        \begin{equation}\label{Eq:FP_epd}
                \begin{aligned}
                        & \epsilon_{\parallel} = \frac{4}{105} {(\frac{B_{\parallel}}{B_{\rm dip}})}^2 \epsilon_{\rm dip} \approx 5.7 \times 10^{-7} B_{\parallel, 16}^2 R_{\rm ns,6}^3 {(\frac{M_{\rm ns}}{1.4 \ M_{\odot}})}^{-1}, \\
                        & \epsilon_{\delta} = \frac{16}{945} {(\frac{B_{\delta}}{B_{\rm dip}})}^2 \epsilon_{\rm dip} \approx 2.5 \times 10^{-7} B_{\delta,16}^2 R_{\rm ns,6}^3 {(\frac{M_{\rm ns}}{1.4 \ M_{\odot}})}^{-1}.
                \end{aligned}
\end{equation}
Combined with Eq.~(\ref{Eq:constr_FP_Bdip}) and Eq.~(\ref{Eq:FP_epd}), the limitation of these two components is
\begin{equation}
        \begin{aligned}
                & B_{\rm \parallel} = \frac{35 T}{R_{\rm ns} {\rm cos}\theta} \sqrt{\frac{5 \pi^3 c M_{\rm ns}}{6 P_{\rm pre}^3 |\Delta \phi|}} \gtrsim 1.2 \times 10^{17} \ {\rm G} \ {({\rm cos} \theta)}^{-1}, \\
                & B_{\rm \delta} = \frac{315 T}{2 R_{\rm ns} {\rm cos}\theta} \sqrt{\frac{5 \pi^3 c M_{\rm ns}}{24 P_{\rm pre}^3 |\Delta \phi|}} \gtrsim 2.8 \times 10^{17} \ {\rm G} \ {({\rm cos} \theta)}^{-1}.
        \end{aligned}
\end{equation}
Even if the ${({\rm cos} \theta)}^{-1}$ takes the minimum value of $1$, it also far exceeds the magnetic field range of the observation limit.

In summary, the free precession model with the ellipticity caused by the internal magnetic field can be adjusted to meet the observation by changing the values of the parameters, and the other two do not fit the observation.

\subsubsection{Constraints on the radiation-driven precession model}

For the radiation-driven precession model, if the magnetic field of the neutron star is represented by the magnetic field of the magnetic dipole, the neutron star is a dipole that is rotating with an angle $\theta$  for the distant observer. The change rate of radiation energy is
\begin{equation}
        \dot{E}_{\rm rot} = \frac{2 \mu^2 \Omega_{\rm rot}^4 {\rm sin}^2 \theta}{3 c^3},
\end{equation}
where $\mu = \frac{1}{2} B_{\rm dip} R^3$ is the magnetic dipole moment. If we ignore the change of the moment of inertia, the change rate of radiation energy could be
\begin{equation}
        \dot{E}_{\rm rot} = \frac{d (\frac{1}{2} I \Omega_{\rm rot}^2)}{d t} = I \Omega_{\rm rot} \dot{\Omega}_{\rm rot}.
\end{equation}
We can then obtain the equation for slowing down the rotation of a neutron star (\citealt{Ghosh_2007}):
\begin{equation}\label{Eq:dotP_spin}
        \dot{P}_{\rm rot} = \frac{2}{3} {(2\pi)}^2 \frac{\mu^2 {\rm sin}^2 \theta}{I c^3 P_{\rm rot}}.
\end{equation}
\cite{Sob_yanin_2020} gave the equation of the internal magnetic field of a precession neutron star:
        \begin{equation}\label{Eq:RP_Bint}
                \begin{aligned}
                        B_{\rm int} & = c \sqrt{\frac{15 I}{2R_{\rm ns}^5 {\rm cos} \theta_m} \frac{P_{\rm rot}}{P_{\rm pre}} \frac{1}{1 + \lambda R_{\rm ns} / R_{\rm g}}} \\
                        & \approx 7.45 \times 10^{17} \ {\rm G} \ \sqrt{\frac{P_{\rm rot}}{P_{\rm pre}}},
                \end{aligned}
        \end{equation}
where $\theta_{\rm m}$ is magnetic inclination, which is the angle between the rotation axis and the magnetic axis. In the second line of this equation, we use the typical value of the neutron star, and we set ${\rm cos} \theta_m = 1$ and $\lambda = 3$, which corresponds to taking $\beta=1$ in $\epsilon_{\rm mag}$.

\begin{figure}
	\resizebox{\hsize}{!}{\includegraphics{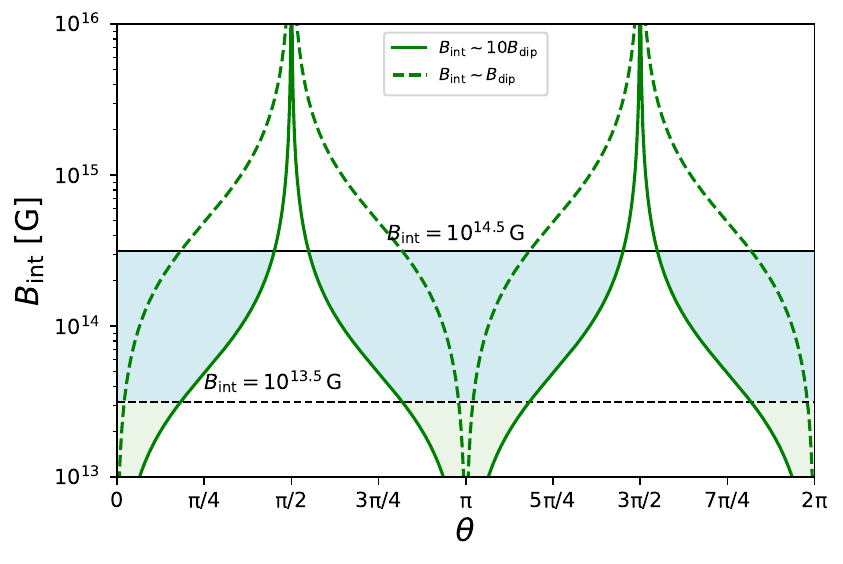}}
    \caption{Relation between the internal magnetic field $B_{\rm int}$ and the angle $\theta$ is drawn according to Eq.~(\ref{Eq:constr_RP_Bint}). We set $\theta = \theta_{\rm m}$ which means the magnetic axis of the neutron star is aligned in the direction of the dipole axis. The black horizontal lines represent the maximum value of the internal magnetic field limited by Eq.~(\ref{Eq:constrain3}). We show two conditions, which are $B_{\rm int}=10^{14.5} \ \rm G$ and $B_{\rm int}=10^{13.5} \ \rm G$. Also, the green lines show the relationship between $B_{\rm int}$ and $\theta$ drawn from Eq.~(\ref{Eq:constr_RP_Bint}). Two different conditions are shown, $B_{\rm int} \sim 10 B_{\rm dip}$ and $B_{\rm int} \sim B_{\rm dip}$. The colored areas show the allowable region.} 
    \label{Fig:constr_RP_Bint-theta}
\end{figure}

Combining this equation with Eq.~(\ref{Eq:dotP_spin}), the change rate of the precession angular velocity is
\begin{equation}
        \begin{aligned}
                \dot{\Omega}_{\rm pre} & = -\frac{30 \pi^3 c G_0^2 M_{\rm ns}^3 B_{\rm dip}^2 {\rm sin}^2 \theta}{R_{\rm ns}^2 {(G_0 M_{\rm ns} + \lambda R_{\rm ns} c^2)}^2 B_{\rm int}^4 P_{\rm pre}^3 {\rm cos}^2 \theta_{\rm m}} \\
                & = (-1.2 \times 10^{-14} \ {\rm s^{-2} \ G^2}) \ \frac{B^2_{\rm dip} {\rm sin}^2 \theta}{B^4_{\rm int} {\rm cos}^2 \theta_{\rm m}}.
        \end{aligned}
\end{equation}
It can be seen that over time, the precession angular velocity of the neutron star will decrease, and the observation period predicted by this model will increase. Combining this equation with Eq.~(\ref{Eq:constrain2}) and $B_{\rm dip} \sim 0.1 B_{\rm int}$, the lower limitation on $B_{\rm int}$ is
\begin{equation}\label{Eq:constr_RP_Bint}
        \begin{aligned}
                B_{\rm int} & = (1.1 \times 10^6 \ \rm{s^{-1} \ G}) \ \frac{T |{\rm sin} \theta|}{|{\rm cos \theta_{\rm m}|} \sqrt{8 |\Delta \phi|}} \\
                & \gtrsim 4.9 \times 10^{13} \ {\rm G} \ \frac{|{\rm sin} \theta|}{|{\rm cos} \theta_{\rm m}|} \sim 4.9 \times 10^{13} \ {\rm G}.
        \end{aligned}
\end{equation}
In the second line of the above equation, we set $\theta = \theta_{\rm m} = \pi / 4$. It is seen that 
if we set $\theta = \theta_{\rm m}$, only $\theta \notin [81.2^{\circ}+k\pi, \ 98.8^{\circ}+k\pi], \ (k \in \mathbb{Z})$ can meet the requirements of the observed magnetic field ($B_{\rm int} \lesssim 10^{14.5} \ \rm G$), as shown in Fig.~\ref{Fig:constr_RP_Bint-theta}.

It should be noted that equation (10) in \cite{Katz_2021} also contains a typo, where the exponent of $P_{\rm pre}$ should be $3$ instead of $1$ and the whole equation is missing a minus sign. The correct equation should be
\begin{equation}
        \dot{\Omega}_{\rm pre}=-\frac{2}{3}{(2\pi)}^3 \frac{R^6 {\rm sin}^2\theta}{I c^3 P_{\rm pre}^3} \frac{{(7.45 \times 10^{17}\ {\rm G})}^4}{{(2B)}^2}.
\end{equation}

In addition, this equation of \cite{Katz_2021} was derived using the approximate formula in the radiation-driven model, and it is more accurate to apply Eq.~(\ref{Eq:constr_RP_Bint}) in this paper.

\subsubsection{Constraints on the geodetic precession model}
For the geodetic precession model, a neuron star with a mass of $M_1$ and a companion with a mass of $M_2 = q M_1$ form a compact binary system. It is assumed that the two stars both orbit around a circle with the same angular velocity in a binary system.

\cite{Yang_Zou_2020} gave the equation of the precession period of a neutron star:
\begin{equation}\label{Eq:GP_Ppre}
                P_{\rm pre} = \frac{1+q}{q(4+3q)} \frac{4 \pi c^2}{G_0 M_1} (1-e^2) a^{5/2},
\end{equation}
where $e$ is the orbital eccentricity and $a$ is the distance between the two stars.

\begin{figure}
	\resizebox{\hsize}{!}{\includegraphics{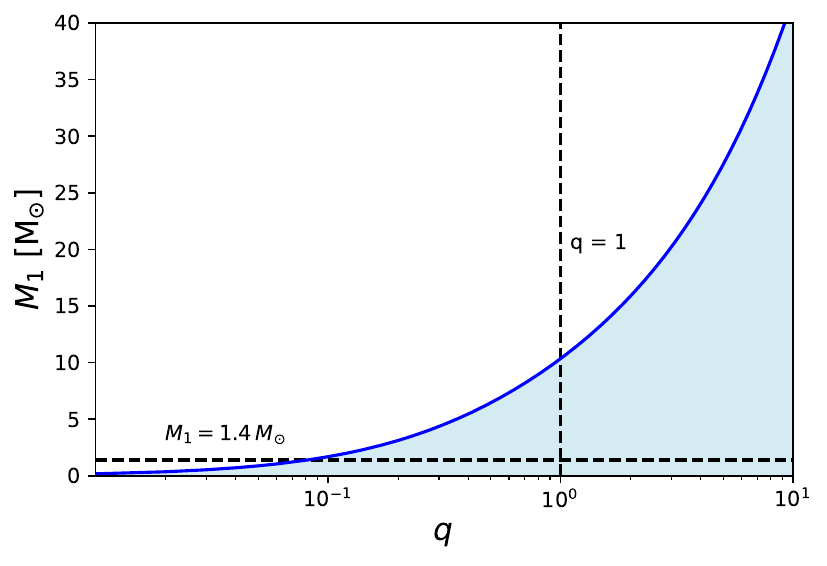}}
    \caption{Relationship between the mass of the neutron star $M_1$ and $q$  drawn according to Eq.~(\ref{Eq:constr_GP_dotOmegapre}), as shown in blue line. The black horizontal solid line represents $M_1 = 1.4 \ M_{\odot}$ and the black vertical solid line shows $q=1$. The blue area of this plot shows the area that matches the observation. }
    \label{Fig:constr_GP_M1-q}
\end{figure}

Neglecting the mass loss and combined with Eq.~(\ref{Eq:GP_Ppre}), the change rate of the precession angular velocity is
\begin{equation}\label{Eq:constr_GP_dotOmegapre}
        \dot{\Omega}_{\rm pre} = 32 \frac{2^{21/5} {\pi}^{13/5} {(1+q)}^{9/5}}{q^{3/5}{(4+3q)}^{8/5}} \frac{{(G_0 M_1)}^{3/5}}{c^{9/5}} P_{\rm pre}^{-13/5}.
\end{equation}
It can be seen that the observation period of the neutron star predicted by this model will decrease over time.  We can obtain the relationship between $q$ and $M_1$  using Eq.~(\ref{Eq:constrain2}) and the 16.3-day observation period of FRB 180916, as shown in Fig.~\ref{Fig:constr_GP_M1-q}. We find that, when $q = 1$, the mass of the neutron star is required to be less than $10 M_{\odot}$, which satisfies the upper limit on the mass of the neutron star; when $M_1 = 1.4 M_{\odot}$, $q \gtrsim 0.1$ is required, which is also consistent with the mass of the compact companion star.

\subsubsection{Constraints on the fall-back disk precession model}


For the fall-back disk precession model with the eccentricity due to the strong magnetic field, \cite{Tong_2020} gave its precession angular velocity as
\begin{equation}\label{Eq:FDP_BtOmegapre}
                \Omega_{\rm pre} = \frac{9 \epsilon}{4} \frac{M_{\theta}}{M_{\rm ns} \kappa^ 3} \Omega_{\rm rot}.
\end{equation}
In this equation, we set $M_\theta = M_0 {\rm cos} \theta_{\rm fb}$ ($M_0$ is the total mass of the fall-back disk and $\theta_{\rm fb}$ is the angle between the axis of rotation of the neutron star and the normal direction of the plane of the fall-back disk), and we set $R = \kappa R_{\rm co}$ ($R$ is the distance between the neutron star and the fall-back disk, $R_{\rm co} = {(G_0 M_{\rm ns} / \Omega_{\rm rot}^2)}^{1/3}$ is the corotation radius, and $R$ is the order of $R_{\rm co}$). In addition, \cite{Tong_2020} thought the value of $M_{\theta}$ is approximately $10^{-6} - 10^{-1} \ M_{\odot}$, and \cite{Qiao_2003_A&A...407L..25Q} assumed $\kappa$ is from $1$ (or smaller) to the value of $2 \sim 3$ when $\theta$ is varied within $0 \sim 90^{\circ}$. We then take the derivative of this equation, and we let $\dot{M}_{\theta} = 0$, obtaining
\begin{equation}\label{Eq:FDP_Bt_dot_Omegapre}
        \dot{\Omega}_{\rm pre} = \frac{9 \epsilon}{4} \frac{M_{\theta}}{M_{\rm ns} \kappa^ 3} \dot{\Omega}_{\rm rot}.
\end{equation}

Assuming that the spin-down of neutron stars is controlled by the magnetic dipole moment, the distance between the neutron star and the fall-back disk remains the same, and $B_{\rm dip} \sim 0.1 B_{\rm int}$,  we can eliminate $\Omega_{\rm rot}$ and $\dot{\Omega}_{\rm rot}$ in Eq.~(\ref{Eq:FDP_Bt_dot_Omegapre}) by combining with Eq.~(\ref{Eq:dotP_spin}). We then can obtain the rate of precession angular velocity change:
\begin{equation}\label{Eq:constr_FBD_dotOmegapre}
    \dot{\Omega}_{\rm pre} = - \frac{640 \pi^3 \mu^2 {\rm sin}^2 \theta M_{\rm ns} \kappa^6}{243 c^3 \epsilon^2 R_{\rm ns}^2 P_{\rm pre}^3 M_{\theta}^2}.
\end{equation}
For $\mu = \frac{1}{2} B_{\rm dip} R_{\rm ns}^3$ and $\epsilon = 10^{-4} B_{\rm int, 16}^2$, we can obtain the internal magnetic field by combining with Eq.~(\ref{Eq:constrain2}):
\begin{equation}
        \begin{aligned}
                B_{\rm int} & = (\frac{4}{9} \times 10^{35} \ \rm{G^2}) \sqrt{\frac{10 \pi^3 \kappa^6 R_{\rm ns}^4 M_{\rm ns} {\rm sin}^2 \theta}{3 c^3 M_{\theta}^2 P_{\rm pre}^3 |\dot{\Omega}|}}\\
                & \gtrsim 6.2 \times 10^{14} \ {\rm G} \ |{\rm sin} \theta| \ \frac{0.1 \ M_{\odot}}{M_{\theta}}.
        \end{aligned}
\end{equation}
In the second line of this equation, we set $\kappa = 1$ and $P_{\rm pre} = 16.3 \ \rm d$. In addition, we take the maximum mass of the fall-back disk as $0.1 M_{\odot}$. If the magnetic field strength is required to meet observations, the angle between the dipole axis and the rotation axis needs to satisfy $\theta \in (-30.6^{\circ} + k \pi, \ 30.6^{\circ} + k \pi ), k \in \mathbb{Z}$.

For the eccentricity due to the rotation, according to \cite{Katz_2021}, for the change rate of the precession angular velocity, we can make the qualitative prediction which is $\dot{\Omega}_{\rm pre} \sim - \Omega_{\rm pre} / t_{\rm diss}$, where $t_{\rm diss}$ is the dissipation time of the fall-back disk. According to Eq.~(\ref{Eq:constrain2}), the lower limit on the dissipation time of the fall-back disk is $9.1 \times 10^9 \ \rm s$. However, there is no clear calculation for the dissipation time of the disk. In the future, this model might be tested using this limitation.

\subsubsection{Constraints on the rotation model}

In this section, we consider the model where the spin-down is due to the fallback of the long-time accretion disk. \cite{Beniamini_2020} gave the final rotation period of the magnetar at the magnetic field decay time $\tau_{\rm B} = 880 \ {\rm yr} \ B_{\rm int,16}^{-6/5}$:
\begin{equation}\label{Eq:R_Prot}
        P_{\rm rot} = P_{\rm c} {(\frac{\tau_{\rm B}}{t_{\rm fb}})}^{3 \zeta / 7},
\end{equation}
where $P_{\rm c} = 1.5 \ {\rm ms} \ B_{\rm dip, 15}^{6/7} \dot{M}_{\rm i, -2}^{-3/7} {(\frac{M_{\rm ns}}{1.4 \ M_{\odot}})}^{-5/7}$ is the rotation period of the magnetar in the steady state, $M_{\rm i}$ is the initial mass of the accretion disk, and $t_{\rm fb}$ is the initial fall-back time.

By combining with Eq.~(\ref{Eq:R_Prot}), setting the initial mass of the fall-back disk to $10^{-2} M_{\rm ns}$, and using $\zeta = 2$, $t_{\rm fb} = 10 \ \rm s,$ and $B_{\rm int} \sim 10 B_{\rm dip} = 10^{14.5} \ \rm G$, we can obtain the change rate of angular velocity with the period of $16.3 \ \rm d$:
\begin{equation}\label{Eq:constr_R_dotOmega}
        |\dot{\Omega}| \approx 2.9 \times 10^{-7} \ {\rm s^{-2}} B_{\rm int, 14.5}^{-6/35}.
\end{equation}
We can then obtain the limit on the systemic moment $N = I \dot{\Omega}$:
\begin{equation}
        N \lesssim \frac{8 \Delta\phi}{T^2} I \sim 4.9 \times 10^{29} \ \rm{dyne \ cm}.
\end{equation}
If this is interpreted as the accretion of a neutron star of mass $1.4 \ M_{\odot}$ and radius $10^6 \ \rm cm$, then considering the most simplified case, the corresponding limit on the accretion rate $\dot{M}$ introduced by $N = \frac{d (I \Omega)}{d t} = \frac{3}{5} \sqrt{G_0 R_{\rm ns} M_{\rm ns}} \dot{M}$ is $\lesssim 5.1 \times 10^{13} \ \rm{g \ s^{-1}} \approx 9.6 \times 10^{-13} \ \rm{M_{\odot}}\ {\rm yr}^{-1}$, which is much smaller than the known accretion rate of massive X-ray binaries (NS-OB stars). This model is therefore not representative. 


\subsection{Testing models using the phase shift}\label{sect:constr_3}


\begin{figure*}
    \centering
    \includegraphics[width=17cm]{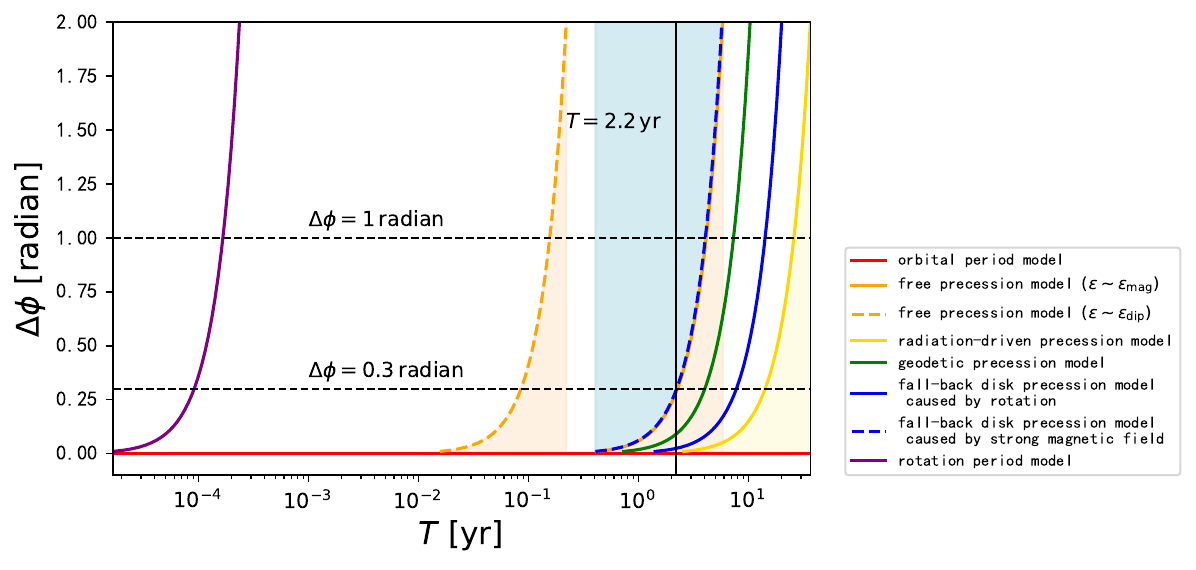}
    \caption{Relation between the phase shift $\Delta \phi$ and the observation duration $T$ for FRB 180916 by combining Eq.~(\ref{Eq:constrain2}) with Eq.~(\ref{Eq:constrain3}) in different models. The orange and yellow lines represent an upper limit and the blue dashed line represents a lower limit. The colored area can fit the observation. The solid vertical line shows $T = 2.2 \ \rm yr$. The rotation period model does not fit the observation of $|\Delta \phi| \lesssim 0.3 \ \rm radian$. The two dashed parallel lines mean $\Delta \phi = 1 \rm \ radian$ and $\Delta \phi = 0.3 \ \rm radian$, respectively. These models can be tested by verifying whether or not there will be a detectable phase shift within a longer observation duration.} 
    \label{Fig:constr_Total_DeltaPhi-T}
\end{figure*}


For the orbital period model, its predicted observation period should be constant. Therefore, regardless of how large $T$ is, $| \Delta \phi |$ is always zero, as shown by the red solid line in Fig.~\ref{Fig:constr_Total_DeltaPhi-T}.

For the free precession model, we can rewrite Eq.~(\ref{Eq:constr_FP_Bdip}) as
\begin{equation}
        \dot{\Omega}_{\rm pre} = - \frac{5 \pi^3 B_{\rm dip}^2 R_{\rm ns}^4}{3 c^3 M_{\rm ns} P_{\rm pre}^3 \epsilon^2 {\rm cos}^2 \theta}.
\end{equation}
If the ellipticity is caused by the internal magnetic field, that is $\epsilon \sim \epsilon_{\rm mag}$, and $B_{\rm int} = 10 B_{\rm dip} \lesssim 10^{14.5} \ \rm G$ is assumed, we obtain
\begin{equation}
        \dot{\Omega}_{\rm pre} = -\frac{\pi^3 G_0^2 M_{\rm ns}^3}{60 c^3 k^2 R_{\rm ns}^4 B_{\rm int}^2 P_{\rm pre}^3 {\rm cos}^2 \theta} \gtrsim -6.7 \times 10^{-18} \ {\rm s^{-2}} \ k^{-2} {({\rm cos} \theta)}^{-2},
\end{equation}
where we set $k \approx 0.12$ and $\theta = 0$. We therefore have $|\dot{\Omega}_{\rm pre}| \lesssim 6.7 \times 10^{-16} \ \rm s^{-2}$. Subsequently, when the phase shift $|\Delta \phi| = 1 ~\rm radian$, the lower limit on the desired duration is approximately $1.3 \times 10^8 \ \rm s \sim 4.2 \ yr$, which can be used to test this model, shown as the orange solid line in Fig.~\ref{Fig:constr_Total_DeltaPhi-T}. If the ellipticity is caused by the dipole field, i.e., $\epsilon \sim \epsilon_{\rm dip}$, we have
\begin{equation}
        \dot{\Omega}_{\rm pre} = - \frac{320 \pi^3 c M_{\rm ns}}{27 R_{\rm ns}^2 B_{\rm dip}^2 {\rm cos}^2 \theta P_{\rm pre}^3} \gtrsim - 3.3 \times 10^{-13} \ \rm s^{-2}.
\end{equation}
Therefore, we have $|\dot{\Omega}_{\rm pre}| \lesssim 3.3 \times 10^{-13} \ \rm s^{-2}$, and if we set $|\Delta \phi| = 1 \ \rm radian$, the lower limit on $T$ approximately is $4.9 \times 10^6 \ \rm s \sim 0.4 \ \rm yr$, as shown by the orange dashed line in Fig.~\ref{Fig:constr_Total_DeltaPhi-T}. But no such phase shift has been observed with current observation time ($T \approx 2.2 \ \rm yr$).

For the radiation-driven precession model, assuming $B_{\rm int} = 10 B_{\rm dip} \lesssim 10^{14.5} \ \rm G$ and combining with Eq.~(\ref{Eq:constr_RP_Bint}), the limit on the change rate of precession angular velocity is
\begin{equation}
        \dot{\Omega}_{\rm pre} \gtrsim -1.2 \times 10^{-17} \rm \ s^{-2}.
\end{equation}
We therefore have $|\dot{\Omega}_{\rm pre}| \lesssim 1.2 \times 10^{-17} \ \rm s^{-2}$, and when $|\Delta \phi| = 1 \ \rm radian$, the lower limit on $T$ is approximately $8.2 \times 10^8 \ \rm s \sim 25.9 \ \rm yr$, as shown by the solid yellow line in Fig.~\ref{Fig:constr_Total_DeltaPhi-T}.

For the geodetic precession model, combining with Eq.~(\ref{Eq:constr_GP_dotOmegapre}) and assuming $q=1$ and $M_1 = 1.4 \ M_{\odot}$, the change rate of angular velocity is
\begin{equation}
        \dot{\Omega}_{\rm pre} = 1.5 \times 10^{-16} \ \rm s^{-2}.
\end{equation}
For $|\Delta \phi| = 1 \ \rm radian$, we can get $T \approx 2.3 \times 10^8 \ \rm s \sim 7 \ \rm yr$, as shown by the green solid line in Fig.~\ref{Fig:constr_Total_DeltaPhi-T}.

For the fall-back disk precession model, if the eccentricity is caused by the rotation of the neutron star, combining with $\dot{\Omega}_{\rm pre} \sim -\Omega/t_{\rm diss}$ and assuming that $t_{\rm diss} \sim 10^{11} \ \rm s,$ which is the age of supernova remnant (\citealt{Katz_2021}), we obtain
\begin{equation}
        |\dot{\Omega}_{\rm pre}| \sim 4 \times 10^{-17} \ \rm s^{-2}.
\end{equation}
Therefore, when $|\Delta \phi| = 1 \rm \ radian$, we obtain $T \approx 15 \ \rm yr$, as shown by the solid blue line in Fig.~\ref{Fig:constr_Total_DeltaPhi-T}. If the eccentricity is caused by the strong magnetic filed, according to Eq.~(\ref{Eq:constr_FBD_dotOmegapre}) and assuming that $\theta = 30.6^{\circ}$, we obtain a lower limit on its change rate of angular velocity of
\begin{equation}
        \dot{\Omega}_{\rm pre} \gtrsim -4.88 \times 10^{-16} \ \rm s^{-2}.
\end{equation}
For $|\Delta \phi| = 1 \ \rm radian$, the upper limit on $T$ is about $1.28 \times 10^8 \ \rm s \sim 4.06 \ \rm yr$, which is larger than the recently observed duration of $2.2 \ \rm yr$, as shown by the dashed blue line in Fig.~\ref{Fig:constr_Total_DeltaPhi-T}. 

For the rotation period model, according to Eq.~(\ref{Eq:constr_R_dotOmega}), its change rate of angular velocity is
\begin{equation}
        \dot{\Omega}_{\rm pre} \sim 2.9 \times 10^{-7} \rm \ s^{-2}.
\end{equation}
For $|\Delta \phi| = 1 \ {\rm radian}$, the observation duration should be $T \approx 5.3 \times 10^3 \ \rm s$, as shown by the purple solid line in Fig.~\ref{Fig:constr_Total_DeltaPhi-T}.

In summary, the above calculations show that the rotation period model cannot fit the observations, but other models could. For the other models, we can test them using further observations. If we can detect a clear phase shift $\Delta \phi$, the orbital period model and the free precession model caused by the dipole field cannot fit the observation. If the phase shift is not detected in the observation duration of 4.06 years, the fall-back disk precession model due to the strong magnetic field is excluded. If the phase shift can be detected when $T$ is smaller than 4.2 years, the free precession model caused by the internal magnetic field does not fit the observation. If $\Delta \phi$ cannot be detected during the observation of 7 years, the geodetic precession model is excluded. If $\Delta \phi$ cannot be detected during a $T$ of 15 years, the fall-back disk precession model does not fit the observations. If $\Delta \phi$ can be detected in $T<25.9 \ {\rm yr}$, the radiation-driven model will be excluded.

\subsection{Constraining the model by comparing the predicted $P_{\rm rot}$ with observed $P_{\rm rot}$}

According to the third observation, the magnetic field should be less than $10^{13.5} - 10^{14.5} \ \rm G$. However, according to \cite{Beniamini_2020}, the rotation period model needs $B_{\rm int} \gtrsim 10^{16} \ \rm G$, and so the rotation period model is not consistent with the third observation.

\begin{figure}
	\resizebox{\hsize}{!}{\includegraphics{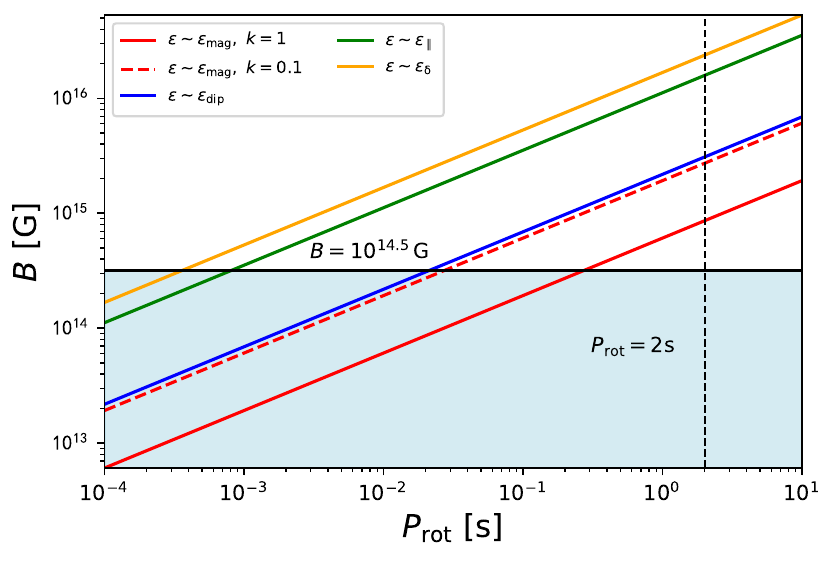}}
    \caption{Relation between the rotation period $P_{\rm rot}$ and the magnetic field $B$ introduced in each free precession model. The vertical dashed line shows $P_{\rm rot} = 2 \ \rm s$. It is seen that when $P_{\rm rot} = 2 \ \rm s$, only the model caused by the internal magnetic field can fit the observation of $B_{\rm int} \lesssim 10^{13.5}-10^{14.5} \ \rm G$ by adjusting the value of $k$.} 
    \label{Fig:constr_B-Prot}
\end{figure}

In the following, we use the third and fourth observations to test these period models (except for orbital period model and geodetic precession model) by discussing the magnetar rotation period. For the free precession model, combining Eqs.~(\ref{Eq:FP_1})-(\ref{Eq:FP_epd}), we can obtain the rotation period of the magnetar, as shown in Fig.~\ref{Fig:constr_B-Prot}. Only the model caused by the internal magnetic field can explain the observation. In addition, even in the model caused by the internal magnetic field, when $P_{\rm rot} = 2 \ \rm s$, to fit the observation of $B_{\rm int} \lesssim 10^{13.5}-10^{14.5} \ \rm G$, the value of $k$ should be larger than $7.4$ which contradicts $k \ll 1$. Therefore, the free precession model is not consistent with observations.

For the radiation-driven precession model, according to Eq.~(\ref{Eq:RP_Bint}), when $B_{\rm int} \lesssim 10^{14.5} \ \rm G$, the rotation period of the magnetar of FRB 180916 should satisfy $P_{\rm rot} \lesssim 0.25 \ \rm s$, which is not consistent with the fourth observation of $P_{\rm rot} = 2 \ \rm s$.

For the fall-back disk precession model, if the eccentricity is caused by the strong magnetic field, according to Eq.(\ref{Eq:FDP_BtOmegapre}), we find that the corresponding rotation period needs to satisfy $P_{\rm rot} \lesssim 0.02 \ \rm s \ \frac{M_{\theta}}{0.1 \ M_{\odot}}$. If we set $P_{\rm rot} = 2 \ \rm s$, the initial mass of the fall-back disk should satisfy $M_{\theta} \gtrsim 10 \ M_{\odot}$, which is not consistent with range of $10^{-6} - 10^{-1} \ M_{\odot}$.

In summary, for FRB 180916, the free precession model, the radiation-driven precession model, the fall-back disk precession model caused by the strong magnetic field, and the rotation period model are not consistent with these two observations.

\section{Summary and Discussion}
\label{sect:sum}

In this paper, we use some recent observations to test several period models for FRB 180916. The geodetic precession model is the most likely periodic model for FRB 180916.
The periodic models can also be tested by comparing the difference in the rate of change of the observed period predicted by various models. For the orbital period model, the observed period will be constant. For the free precession model, the radiation-driven precession model, and the rotation period model, the observed period will increase over time. For the geodetic precession model, the observed period will decrease over time. For the fall-back disk precession model, the observed period will increase or decrease over time. Therefore, in the future, we can further test the geodetic precession model by using this characteristic.

Moreover, if we do not consider the fourth observation of Sect.~\ref{sect:obser}, the free precession model caused by internal magnetic field, the radiation-driven precession model, the geodetic precession model, and the fall-back disk precession model can fit other observations. We can further test these models using the observed phase shift over longer observation times in the future. 
With increasing observation times and data, several periodic models introduced in this paper can be further limited using the above methods. If more periodic FRBs are found, their models can also be tested by imitating the above process.

\begin{acknowledgements}
We thank the anonymous referee for helpful comments. This work was supported by the National Natural Science Foundation of China (grant No. U1831207), and the Fundamental Research Funds for the Central Universities (No. 0201-14380045) 

\end{acknowledgements}
  
\bibliographystyle{aa}
\bibliography{ref_frb}

\end{document}